\documentclass[useAMS,usenatbib]{mn2e}
\usepackage[dvips]{graphicx}

\title[Modelling of c-C$_{2}$H$_{4}$O formation on grain surfaces]
  {Modelling of c-C$_{2}$H$_{4}$O formation on grain surfaces}
\author[A. Occhiogrosso et al]
  {A.~Occhiogrosso,$^1$\thanks{E-mail: ao@star.ucl.ac.uk.}
  S. Viti,$^1$ M. D. Ward,$^2$ and S. D. Price$^2$ \\
  $^1$Dept. of Physics and Astronomy, UCL, Gower Place, London WC1E6BT, UK \\
  $^2$Dept. of Chemistry, UCL, 20 Gordon Street, London, WC1H 0AJ }
\date{Released 2012 Xxxxx XX}

\pagerange{\pageref{firstpage}--\pageref{lastpage}} \pubyear{2012}

\def\LaTeX{L\kern-.36em\raise.3ex\hbox{a}\kern-.15em
    T\kern-.1667em\lower.7ex\hbox{E}\kern-.125emX}

\begin{document}

\label{firstpage}

\maketitle

\begin{abstract}
Despite its potential reactivity due to ring strain, ethylene oxide (c-C$_{2}$H$_{4}$O) is a complex molecule that seems to be stable under the physical conditions of an interstellar dense core; indeed it has been detected towards several high-mass star forming regions with a column density of the order of 10$^{13}$cm$^{-2}$ (Ikeda et al. 2001). To date, its observational abundances cannot be reproduced by chemical models and this may be due to the significant contribution played by its chemistry on grain surfaces.
Recently, Ward \& Price (2011) have performed experiments in order to investigate the surface formation of ethylene oxide starting with oxygen atoms and ethylene ice as reactants.
We present a chemical model which includes the most recent experimental results from Ward \& Price (2011) on the formation of c-C$_{2}$H$_{4}$O. We study the influence of the physical parameters of dense cores on the abundances of c-C$_{2}$H$_{4}$O. We verify that ethylene oxide can indeed be formed during the cold phase (when the ISM dense cores are formed), via addition of an oxygen atom across the C=C double bond of the ethylene molecule, and released by thermal desorption during the hot core phase. A qualitative comparison between our theoretical results and those from the observations shows that we are able to reproduce the abundances of ethylene oxide towards high-mass star-forming regions.

\end{abstract}


\section{Introduction}
Ethylene oxide (c-C$_{2}$H$_{4}$O) is the simplest epoxide, a ring-shaped organic compound involving an oxygen atom bonded to two carbon atoms.
Since it is believed to lead to the synthesis of amino acids (Cleaves 2003) and hence potentially influence early pre-biotic pathways (Miller \& Schlesinger 1993), c-C$_{2}$H$_{4}$O may play an important role in the interstellar medium (ISM).

First detected towards the Sgr B2N molecular cloud associated with 10 rotational transitions (Dickens et al. 1997), this molecule has been investigated in a more extended frequency range (662 lines between 262-358 GHz)(Pan et al. 1998) and more recently Bernstein \& Lynch (2009) attributed the unidentified IR band features discovered by Gillett et al. (1973) to ethylene oxide and to another small carbonaceous molecule, cyclopropenylidene. Ikeda et al. (2001) estimated the c-C$_{2}$H$_{4}$O fractional abundances by observing 20 massive star-forming regions and two dark clouds. In fact, ethylene oxide was detected towards 10 of these sources but it was not found in TMC-1 and TMC-1(NH$_{3}$) dark clouds, in contrast to one of its isomers, acetaldehyde (CH$_{3}$CHO) which seems to be ubiquitous. Ikeda et al. (2001) also investigated the relationship between c-C$_{2}$H$_{4}$O abundances and the dust temperature, but without finding any correlation. On the other hand, the evidence of similar trends in the column densities of all C$_{2}$H$_{4}$O isomers and ethanol identified the latter as a possible precursor of ethylene oxide and its isomers. Indeed, a gas-phase pathway for ethylene oxide formation has been proposed (Dickens et al. 1997) with ethanol as one of the reactants:

CH$_{3}$$^{+}$ + C$_{2}$H$_{5}$OH $\rightarrow$ C$_{2}$H$_{5}$O$^{+}$ + CH$_{4}$

C$_{2}$H$_{5}$O$^{+}$ + e$^{-}$ $\rightarrow$ C$_{2}$H$_{4}$O + H

where the formula C$_{2}$H$_{4}$O does not distinguish among the different isomers.
However, Ikeda et al. (2001) performed simulations of c-C$_{2}$H$_{4}$O chemical evolution by including only pure gas-phase reactions at different temperatures and volume densities and they found that the calculated abundances of ethylene oxide decreased when the gas-kinetic temperature was raised. The results of Ikeda et al. (2001) suggested that other reactions such as grain surface reactions may be important for the synthesis of ethylene oxide.
  
Another gas-phase pathway for the formation of ethylene oxide has been proposed by Turner \& Apponi (2001) which studied the reaction chain to produce c-C$_{2}$H$_{4}$O and one of its isomers, vinyl alcohol (CH$_{2}$CHOH), starting with the interaction between CH$_{3}$$^{+}$ and H$_{2}$CO; unfortunately, none of the intermediates of the latter reaction have been identified yet in the ISM. The abundance ratios among the three isomers, acetaldehyde, ethylene oxide and vinyl alcohol, seem to be very uncertain (Bennet et al. 2005), but generally CH$_{3}$CHO was found to be the most abundant and CH$_{2}$CHOH has been predicted to have a lower abundance than ethylene oxide in the experiments by Bennet et al. (2005). This comparison could be explained in the light of the fact that the formation of ethylene oxide occurs via direct addition of the oxygen across the C-C double bond of ethylene, while vinyl alcohol requires an insertion of an oxygen into the C-H bond. Hudson \& Moore (2003) suggested a solid state route for the formation of CH$_{2}$CHOH involving exposure of H$_{2}$O:C$_{2}$H$_{2}$ ices to protons and photons, but the first proposed surface scheme for ethylene oxide formation dates back to Charnley (2004) who identified acetaldehyde (CH$_{3}$CHO) as a possible precursor.
The contribution of grain-surface reactions to the production of the three isomers with formula  C$_{2}$H$_{4}$O has also been investigated by Bennet et al. (2005): they presented the results of electron irradiation on a mix of carbon dioxide and ethylene ice, under interstellar and cometary conditions, to highlight the effect of non-thermal processing of ices.

Recently, an alternative solid state pathway for the formation of ethylene oxide has been investigated by Ward \& Price (2011) who studied the surface reactivity between C$_{2}$H$_{4}$ and thermal O atoms under astrophysically relevant conditions. They co-deposited the reactants on a graphite surface at various temperatures in the 12-90 K range. In order to identify the products formed, they recorded desorption spectra and they also selectively ionized the three different isomers with a tunable dye laser. Finally, they performed a similar laboratory study with C$_{3}$H$_{6}$ as reactant, but unfortunately although C$_{3}$H$_{6}$O was observed they were unable to identify the isomeric products of the reaction due to the large number of possible molecules with formula C$_{3}$H$_{6}$O. Despite these limitations, the authors assumed that the majority of the C$_{3}$H$_{6}$O formed is propylene oxide by following the same mechanism as for the case of ethylene oxide production. Unlike c-C$_{2}$H$_{4}$O, c-C$_{3}$H$_{6}$O has not yet been detected in the ISM since it is believed to isomerize to propanal or acetone, which have both been observed in the ISM (Hollis et al. 2004 and Combes et al. 1987, Snyder et al. 2002).
To date, the observed abundances of ethylene oxide have not yet been reproduced by chemical models; in fact, to our knowledge, none of the grain-surface pathways for formation for ethylene oxide have been included in the models that simulate the star formation process.
  
In this paper, we update the UCL\_CHEM gas-grain chemical model with the reaction rates for the production of this species as evaluated by Ward \& Price (2011). By following the same logic as these authors, we extended our study to include the formation of propylene oxide in order to verify if C$_{3}$H$_{6}$O can be produced under ISM conditions even without having any observational evidence of its presence in the interstellar space. 
The paper is organized as follows: Section 2 gives details of the model; Section 3 considers the astronomical implications of our study; Section 4 contains a discussion of our results and it also shows a comparison between our theoretical column densities and those taken from the observations. Finally, we give our conclusions in Section 5. 

\section{The Chemical Model}
UCL\_CHEM is a time-depth dependent gas-grain chemical model which simulates the abundances of molecules in the ISM and star-forming regions.
The first version of the code was developed by Viti \& Williams (1999) and revisited by the same authors in 2004 (Viti et al. 2004). Our model has been extensively described in the literature (see Occhiogrosso et al. 2011 for more details) so here we only highlight the main points.

The process of star formation is simulated in two steps: during Phase I the gas collapses to a dense core from a diffuse and atomic state; at this stage the gas freezes onto the  grains and surface reactions occur.
During Phase II, the gas and the dust surrounding the central star heat up, causing the sublimation of the icy mantles. The thermal desorption can be set up to occur instantaneously or as a function of time as in Viti et al. (2004). In the latter case, molecules are classified in five different categories as described by the experimental study of Collings et al. (2004): (i) CO-like; (ii) H$_{2}$O-like; (iii) intermediate; (iv) reactive and (v) refractory. In the present work, we consider c-C$_{2}$H$_{4}$O as H$_{2}$O-like (it will co-desorb when the H$_{2}$O ice desorbs) based on the fact that in the Ward \& Price (2011) experiments, its desorption profile shows only one peak (as for the water desorption profile) and its dipole moment is big enough to assume that the species will have a lower desorption energy from the surface (in that specific case, graphite) than from itself. In our model the temperature of the gas is derived as a function of the luminosity (and therefore the age) of the accreting protostar through a simple power law (see Viti et al. 2004 for details).
The model includes two networks: the gas-phase network is based on the UMIST database (Woodall et al. 2000) (even if some coefficients have been modified with those from the most recent database KIDA (Wakelam et al. 2009)) with some updates discussed later (see Section 3), and a grain-surface network with reactions whose rates are experimentally determined where possible.

For this study, we include in our surface reaction network the formation of c-C$_{2}$H$_{4}$O and c-C$_{3}$H$_{6}$O on the grain surface as investigated by Ward \& Price (2011) (see Table 1). In order to analyze their data, these authors developed a simple kinetic model in which molecules can react via Eley-Rideal (ER) or Langmuir-Hinshelwood (LH) mechanisms. In the latter case, both reactants adsorb and thermalize on the surface before a diffusion reaction occurs. We consider the reactions to occur via the LH mechanism that seems of greater astrophysical relevance as stated by Awad et al. (2005) and by the authors themselves (see their paper for the details).

\begin{table}
 \begin{minipage}{126mm}
  \caption{Surface reactions investigated by Ward \& Price (2011)}
  \label{anymode}
  \begin{tabular}{@{}|cc}
  \hline
$^{a}$Reaction & $\alpha$ (cm$^{2}$mol$^{-1}$s$^{-1}$) at 20K\\
   \hline
mO + mC$_{2}$H$_{4}$ $\rightarrow$ mc-C$_{2}$H$_{4}$O & 1.20x10$^{-19}$\\

mO + mC$_{3}$H$_{6}$ $\rightarrow$ mc-C$_{3}$H$_{6}$O & 3.40x10$^{-19}$\\
 \hline
  \end{tabular}
\footnotetext[1]{The m before the molecular formul$\ae$ stands for $mantle$}
 \end{minipage}
\end{table}
 
Since our code accounts for reactions that occur in three dimensions (i.e. gas-phase reactions), we need to transform the surface rate constants of Ward and Price discussed above. Therefore we first evaluate two quantities: the volume of ice per grain and the number density of grains. In the latter case we assume a dust/mass ratio of 0.01 and a dust density of 2.5 g cm$^{-3}$. Thus, we obtain a grain number density of 3.8x10$^{-5}$ cm$^{-3}$. In order to estimate the volume of ice we consider the grains as spheres and we assume a thick ice layer of 0.3 $\micron$ (Collings et al. 2003). Since the major partner in the ice mixture is water we also account for the fraction of C$_{2}$H$_{4}$ or C$_{3}$H$_{6}$ ice compared to water ice in order to avoid overestimating our rate. We consider that not all the oxygen ice will react with the ethylene or propene ice because of the presence of other ice components in between. We hence estimate a final value of 6x10$^{-15}$ cm$^{3}$s$^{-1}$ at 20 K for the reaction between O and C$_{2}$H$_{4}$ and a rate of 5x10$^{-14}$cm$^{3}$s$^{-1}$ at 20 K for the reaction between O and C$_{3}$H$_{6}$ onto the grains.

In order to reveal the contribution of these rates to the abundance of c-C$_{2}$H$_{4}$O and c-C$_{3}$H$_{6}$O in the ISM, we initially did not include any gas-phase pathway for the formation of these species.

The percentage of depletion of oxygen atoms on the grains is certainly one of the critical parameters in the formation of ethylene and propylene oxide; Caux et al. (1999) found that the fractional abundance of oxygen (relative to hydrogen nuclei) on the grains is of the order of 10$^{-6}$ and this amount depends on the percentage of the oxygen that is depleted in the dense, cold cores. In our model, atomic oxygen freezes on the grains in the form of H$_{2}$O, O and OH. It is known that oxygen containing species in interstellar ices consist mainly of H$_{2}$O and also that the fractional abundances of water ice (relative to the total number of hydrogen nuclei) are of the order of 10$^{-4}$ (Caux et al. 1999); we vary the percentage contribution of each channel by which oxygen containing species accumulate on the grains by varying the efficiency of the different oxygen accretion pathways as shown in Table 2. 
 
\begin{table}
 \begin{minipage}{126mm}
  \caption{List of test chemical models.}
  \label{symbol}
  \begin{tabular}{@{}|cccc}
  \hline
Test & mH$_{2}$O (\%) & mO (\%) & mOH (\%)\\
   \hline
Test1 & 99.5 & 0.5 & 0 \\
Test2 & 90 & 1 & 9 \\
Test3 & 90 & 2 & 8 \\
Test4 & 90 & 5 & 5 \\
Test5 & 90 & 10 & 0 \\
Test6 & 80 & 20 & 0 \\
Test7 & 70 & 15 & 15 \\
   \hline
  \end{tabular}
 \end{minipage}
\end{table}

Since we find 1 \% of oxygen atoms in the ice (mO) is sufficient for the observational values of water ice to be satisfied we use this percentage subsequently for our model. 
Figure 1 shows the fractional abundances of our selected species as a function of the time at the end of Phase I for a model with a final density of 10$^{7}$ cm$^{-3}$. During this phase atoms and molecules freeze onto the grains and surface reactions can occur due to the catalytic effects of the icy mantles. The dotted and dashed lines refer to the reactants, atomic oxygen and ethylene respectively, the solid line shows the ethylene oxide formation. Note that the fractional abundance of ethylene at the beginning of Phase I is low since it also mainly forms on the grains.
We mostly focus on ethylene oxide because results show that c-C$_{3}$H$_{6}$O is produced below the threshold of detection; this fact may mean that this molecule is not detectable in the ISM although we stress that we do not have a complete network of reactions (both in gas-phase and on grain surfaces) for c-C$_{3}$H$_{6}$O and its precursors. 

\begin{figure}
 \includegraphics[width=60mm, angle=270]{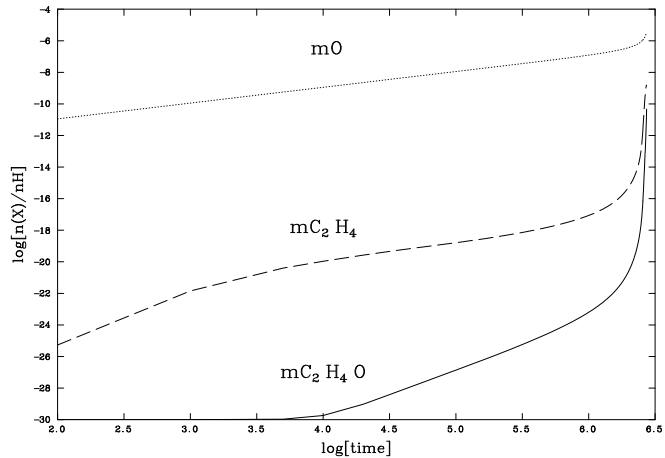}
 \caption{Grain surface formation of selected species during Phase I. The fractional abundances (with respect to the total number of hydrogen nuclei) vary along with the time.}
\end{figure}

We conclude that the reaction between oxygen and ethylene on grains with a reaction rate of 6x10$^{-15}$ cm$^{3}$s$^{-1}$ (evaluated as described above) may be the main formation path for c-C$_{2}$H$_{4}$O.
 
In order to compare our theoretical results with observations we ran further models where gas-phase reactions were included despite the fact that the routes proposed to date (as mentioned in the Introduction) do not distinguish among the different isomers with formul$\ae$ C$_{2}$H$_{4}$O.
We then include in our model the gas-phase routes of formation for c-C$_{2}$H$_{4}$O listed in Table 3. We insert the C$_{2}$H$_{4}$O species in the UMIST database with the reaction rates taken from the OSU database (http://www.physics.ohio-state.edu/~eric/research.html). Since the literature does not distinguish among the different isomers, we assume the ratios of 2:1.5:1 (as reported by Ikeda et al. 2001) for acetaldehyde, ethylene oxide and vinyl alcohol respectively to evaluate the rate of formation of each product.

\begin{table*}
 \begin{minipage}{126mm}
  \caption{Gas-phase paths of c-C$_{2}$H$_{4}$O formation and destruction. The rates are taken from the OSU database and they have been scaled by considering the ratio 2:1.5:1 as in Ikeda et al. (2001). R indicates a reactant, P a product.}
  \label{symbol}
  \begin{tabular}{@{}|cccccc}
   \hline
R1 & R2 & P1 & P2 & P3 & $k$$\backslash$(cm$^{3}$s$^{-1}$)\\
   \hline
C$^{+}$ & C$_{2}$H$_{4}$O & C$_{2}$H$_{3}$O$^{+}$ & CH & & 7.3x10$^{-9}$\\
C$^{+}$ & C$_{2}$H$_{4}$O & C$_{2}$H$_{4}$O$^{+}$ & C & & 7.3x10$^{-9}$\\
H$^{+}$ & C$_{2}$H$_{4}$O & C$_{2}$H$_{3}$O$^{+}$ & H$_{2}$ & & 2.2x10$^{-10}$\\
H$^{+}$ & C$_{2}$H$_{4}$O & C$_{2}$H$_{4}$O$^{+}$ & H & & 2.2x10$^{-10}$\\
He$^{+}$ & C$_{2}$H$_{4}$O & HCO$^{+}$ & CH$_{3}$ & He & 6.1x10$^{-9}$\\
He$^{+}$ & C$_{2}$H$_{4}$O & CH$_{3}$$^{+}$ & HCO & He & 6.1x10$^{-9}$\\
He$^{+}$ & C$_{2}$H$_{4}$O & C$_{2}$H$_{2}$O$^{+}$ & H$_{2}$ & He & 6.1x10$^{-9}$\\
He$^{+}$ & C$_{2}$H$_{4}$O & C$_{2}$H$_{3}$O$^{+}$ & H & He & 6.1x10$^{-9}$\\
He$^{+}$ & C$_{2}$H$_{5}$OH & C$_{2}$H$_{3}$O$^{+}$ & H$_{2}$ & & 2.9x10$^{-9}$\\
S$_{2}$$^{+}$ & C$_{2}$H$_{5}$OH & H$_{2}$S$_{2}$$^{+}$ & C$_{2}$H$_{4}$O & & 6.4x10$^{-11}$\\
H$_{3}$$^{+}$ & C$_{2}$H$_{4}$O & C$_{2}$H$_{5}$O$^{+}$ & H$_{2}$ & & 2.7x10$^{-8}$\\
HCO$^{+}$ & C$_{2}$H$_{4}$O & C$_{2}$H$_{5}$O$^{+}$ & CO & & 1.1x10$^{-8}$\\
H$_{3}$O$^{+}$ & C$_{2}$H$_{4}$O & C$_{2}$H$_{5}$O$^{+}$ & H$_{2}$O & & 1.2x10$^{-8}$\\
O & C$_{2}$H$_{5}$$^{+}$ & C$_{2}$H$_{4}$O$^{+}$ & H & & 7.5x10$^{-11}$\\
CH$_{3}$$^{+}$ & C$_{2}$H$_{4}$O & C$_{3}$H$_{6}$OH$^{+}$ & & & 4.3x10$^{-10}$\\
e$^{-}$ & C$_{2}$H$_{4}$O$^{+}$ & CH$_{3}$ & HCO & & 6.5x10$^{-7}$\\
e$^{-}$ & C$_{2}$H$_{4}$O$^{+}$ & C$_{2}$H$_{2}$O & H & H & 6.5x10$^{-7}$\\
e$^{-}$ & C$_{2}$H$_{4}$O$^{+}$ & C$_{2}$H$_{2}$O & H$_{2}$ & & 6.5x10$^{-7}$\\
e$^{-}$ & C$_{2}$H$_{5}$O$^{+}$ & C$_{2}$H$_{4}$O & H & & 6.5x10$^{-7}$\\
e$^{-}$ & C$_{2}$H$_{5}$OH$^{+}$ & C$_{2}$H$_{4}$O & H$_{2}$ & & 6.5x10$^{-7}$\\
e$^{-}$ & C$_{2}$H$_{5}$OH$_{2}$$^{+}$ & C$_{2}$H$_{4}$O & H$_{2}$ & H & 6.5x10$^{-7}$\\
e$^{-}$ & C$_{2}$H$_{6}$CO$^{+}$ & C$_{2}$H$_{4}$O & CH$_{2}$ & & 6.5x10$^{-7}$\\
e$^{-}$ & C$_{3}$H$_{6}$OH$^{+}$ & C$_{2}$H$_{4}$O & CH$_{3}$ & & 6.5x10$^{-7}$\\
C$_{2}$H$_{4}$O & photon & CH$_{3}$ & HCO & & 2.2x10$^{-10}$\\
C$_{2}$H$_{4}$O & photon & CH$_{4}$ & CO & & 2.2x10$^{-10}$\\
C$_{2}$H$_{4}$O & photon & C$_{2}$H$_{4}$O$^{+}$ & e$^{-}$ & & 1.5x10$^{-10}$\\
   \hline
  \end{tabular}
 \end{minipage}
\end{table*}

The different paths listed in Table 3 are mostly two-body reactions. Specifically we have charge exchange and ion-neutral reactions, electronic recombination of positive ions with electrons followed by the dissociation of the resulting molecule (dissociative attachment) and finally photo-processes where the dissociation or ionization of neutral species is performed by UV photons from a standard interstellar UV field. In the latter case, the reaction rates depend on the visual extinction of the source so we do not expect it to be efficient at the high visual extinction of hot cores and hot corinos.

Before testing the sensitivity of the c-C$_{2}$H$_{4}$O abundances to the different physical conditions of potential astronomical sources, we investigate another important parameter for the formation of ethylene oxide: the amount of C$_{2}$H$_{4}$ available on grain surfaces. Despite the fact that C$_{2}$H$_{4}$ ice has not been detected to date, several experiments (Hiraoka et al. 2000; Kaiser \& Roessler 1998; Barratta et al. 2003; Bennett et al. 2006) have looked at its formation on grains under physical conditions relevant for the interstellar medium. Using different experimental techniques, C$_{2}$H$_{4}$ was observed as a product following irradiation of CH$_{4}$-containing icy mixtures and C$_{2}$H$_{4}$ was also predicted to be an intermediate in the hydrogenation of C$_{2}$H$_{2}$ (Kaiser \& Roessler 1998; Bennett et al. 2006). In these cases, ethylene formation competes with the formation of the C$_{2}$H$_{6}$ that was found the most abundant molecule in the product mixture. In our model we include the hydrogenation of C$_{2}$H$_{2}$ on grains in Phase I to form C$_{2}$H$_{4}$ as well as C$_{2}$H$_{6}$ (see Table 4).

\begin{table}
 \begin{minipage}{80mm}
  \caption{The main accretion and hydrogenation pathways for C$_{2}$H$_{2}$, C$_{2}$H$_{4}$, C$_{2}$H$_{6}$ on grain surfaces.}
  \label{symbol}
  \begin{tabular}{@{}|cc}
  \hline
Reaction & Branching ratios (\%)\\
   \hline
1 C$_{2}$H$_{4}$ $\rightarrow$ mC$_{2}$H$_{4}$ & 10-1 \\
2 C$_{2}$H$_{6}$ $\rightarrow$ mC$_{2}$H$_{6}$ & 100 \\
3 C$_{2}$H$_{2}$ $\rightarrow$ mC$_{2}$H$_{2}$ & 10 \\
4 C$_{2}$H$_{2}$ $\rightarrow$ mC$_{2}$H$_{4}$ & 40-1 \\
5 C$_{2}$H$_{2}$ $\rightarrow$ mC$_{2}$H$_{6}$ & 50-89 \\
6 C$_{2}$H$_{4}$ $\rightarrow$ mC$_{2}$H$_{6}$ & 90-99 \\
   \hline
  \end{tabular}
 \end{minipage}
\end{table}

Based on the experimental studies performed to date (Barratta et al. 2006), each time C$_{2}$H$_{6}$ is produced, a small experimental signature that has been assigned to icy ethylene can also be observed; it is therefore appropriate to investigate the minimum amount of ethylene on grains necessary to reproduce the relevant c-C$_{2}$H$_{4}$O observations. In a similar manner as described above for the case of oxygen, we vary the branching ratios for the different channels involved in the relevant hydrogenation pathways (see Table 4) in order to determine the minimum concentration of C$_{2}$H$_{4}$ on grains needed for the formation of ethylene oxide. As shown in Table 4, we investigated the variation of the branching of C$_{2}$H$_{2}$ hydrogenation to mC$_{2}$H$_{4}$ (m denotes a molecule in an icy mantle) between 1$\%$ and 40$\%$ (Table 4) and we obtain mC$_{2}$H$_{4}$ fractional abundances (relative to the total hydrogen nuclei) of 9.31x10$^{-11}$ and 2.53x10$^{-9}$ respectively. We find the lower abundance of mC$_{2}$H$_{4}$ still generates viable quantities of c-C$_{2}$H$_{4}$O and so we fixed the branching of reaction 4 to 1$\%$ in all further studies. We then investigated the effect of the role of accretion of gas-phase C$_{2}$H$_{4}$ to form mC$_{2}$H$_{4}$, a process which competes with hydrogenation of C$_{2}$H$_{4}$ to yield mC$_{2}$H$_{6}$ (reaction 1 and 6, Table 4). Here we find that a branching of only 1$\%$ in the direction of mC$_{2}$H$_{4}$ still results in a c-C$_{2}$H$_{4}$O abundance in agreement with observations.

In summary, the above investigations of the hydrogenation pathways showed that a branching of only 1$\%$ towards both reactions 1 and 4 in Table 4 yields c-C$_{2}$H$_{4}$O abundances (3.9x10$^{-12}$ cm$^{-2}$) comparable with observations and these branching ratios were used in all further modelling. Finally, we qualitatively model a sample of high-mass star-forming regions by looking at the influence of the physical parameters of the sources on the ethylene oxide abundances. Models are discussed in Section 3.

\section{Astronomical Implications}

Based on the study by Ikeda et al. (2001) ethylene oxide has been detected towards the Orion compact ridge and several hot cores. We qualitatively model these regions with UCL\_CHEM by varying the final sizes and densities of Phase I and the temperatures in Phase II. In Phase I the non-thermal desorption is included as in Roberts et al. (2007). During Phase II icy mantles sublimate as in Viti et al. (2004).
In order to investigate the sensitivity of the chemistry to the physical parameters of dense cores we run a grid of 16 chemical models by varying: (i) the size of the core, (ii) the final density of the collapsing core, (iii) the temperature, (iv) the adsorption energy (controlled by $\epsilon$) and (v) the percentage of the accreted species onto the grain surfaces.
Table 5 lists the different models and the c-C$_{2}$H$_{4}$O fractional abundances ($f$) (relative to hydrogen nuclei) obtained as outputs (see the last column). We also derive the column density (N) by using the formula below:

\begin{equation}
N(c-C_{2}H_{4}O) = X \times A_{\rm v} \times N(H_{2}),
\end{equation}

where X is the ethylene oxide fractional abundance, $A_{\rm v}$ is the visual extinction and N(H$_{2}$) is equal to 1.6x10$^{21}$ cm$^{-2}$ which is the hydrogen column density in 1 magnitude.

\begin{table*}
 \begin{minipage}{126mm}
  \caption{List of chemical models and their parameters: size, density (n$_{H}$), gas temperature (T) during Phase II, threshold adsorption energy ($\epsilon$), efficiency of the freeze-out (fr) during phase I, the percentage of mantle CO (mCO) given by the freeze-out parameter at the end of Phase I of the chemical model. The last column lists the c-C$_{2}$H$_{4}$O theoretical fractional abundances ($f$) (relative to the total number of hydrogen nuclei) obtained as outputs from each model. We also report its calculated column densities (N).}
  \label{symbol}
  \begin{tabular}{@{}|ccccccccc}
  \hline
 & Size (pc) & n$_{H}$/cm$^{-3}$ & T/K & $\epsilon$ & fr & mCO (\%) & N/cm$^{-2}$ & $f$ \\
   \hline
M1 & 0.02 & 2x10$^{8}$ & 300 & 0.1 & 0.3 & 98 & 1.1x10$^{13}$ & 1.8x10$^{-11}$  \\
M2 & 0.02 & 2x10$^{8}$ & 300 & 0.01 & 0.3 & 98 & 6.2x10$^{12}$ & 1.0x10$^{-11}$ \\
M3 & 0.02 & 2x10$^{8}$ & 300 & 0.1 & 0.0075 & 75 & 1.1x10$^{12}$ & 1.8x10$^{-12}$ \\
M4 & 0.02 & 2x10$^{8}$ & 300 & 0.01 & 0.0075 & 75 & 1.1x10$^{12}$ & 1.7x10$^{-12}$\\
M5 & 0.02 & 2x10$^{7}$ & 200 & 0.1 & 0.3 & 98 & 6.8x10$^{12}$ & 1.1x10$^{-11}$\\
M6 & 0.02 & 2x10$^{7}$ & 200 & 0.01 & 0.3 & 98 & 3.9x10$^{12}$ & 6.4x10$^{-12}$\\
M7 & 0.02 & 2x10$^{7}$ & 200 & 0.1 & 0.0075 & 75 & 6.2x10$^{13}$ & 1.0x10$^{-10}$\\
M8 & 0.02 & 2x10$^{7}$ & 200 & 0.01 & 0.0075 & 75 & 6.1x10$^{13}$ & 9.9x10$^{-11}$\\
M9 & 0.04 & 2x10$^{8}$ & 300 & 0.1 & 0.3 & 98 & 2.2x10$^{13}$ & 1.8x10$^{-11}$\\
M10 & 0.04 & 2x10$^{7}$ & 200 & 0.1 & 0.3 & 98 & 1.5x10$^{13}$ & 1.2x10$^{-11}$\\
M11 & 0.06 & 2x10$^{6}$ & 80 & 0.1 & 0.5 & 98 & 2.7x10$^{14}$ & 1.5x10$^{-10}$\\
M12 & 0.06 & 2x10$^{6}$ & 80 & 0.01 & 0.5 & 98 & 1.5x10$^{14}$ & 8.1x10$^{-11}$ \\
M13 & 0.06 & 2x10$^{6}$ & 80 & 0.1 & 0.25 & 75 & 1.5x10$^{14}$ & 8.1x10$^{-11}$\\
M14 & 0.06 & 2x10$^{6}$ & 80 & 0.01 & 0.25 & 75 & 1.6x10$^{14}$ & 8.7x10$^{-11}$\\
M15 & 0.06 & 2x10$^{6}$ & 80 & 0.1 & 0.1 & 40 & 6.8x10$^{14}$ & 3.7x10$^{-10}$\\
M16 & 0.06 & 2x10$^{6}$ & 80 & 0.01 & 0.1 & 40 & 6.4x10$^{14}$ & 3.5x10$^{-10}$\\
   \hline
  \end{tabular}
 \end{minipage}
\end{table*}

\subsection{Sensitivity to variation in the density, temperature and size core}

Models from M1 to M10 reproduce the typical physical parameters for an hot core model as described in Lerate et al. (2008); models M11-M16 best represent the characteristics for a compact ridge where we set a temperature of 80 K in order to look at the molecular abundances when the water desorption has not yet occurred. We consider a star of 25 M$\odot$ and 0.02-0.06 pc in size, characterized by a visual extinction of about 300-500 mags.
We then vary the density of the core (n$_{H}$) in the 10$^{6}$-10$^{8}$ cm$^{-3}$ range and we also modify the temperature (T) according to the observations towards the different astronomical sources, even if for T $\ge$ 100 K we do not expect any influence of this parameter since, as already explained, the sublimation of ethylene oxide occurs simultaneously with the water desorption (at around 100 K). 

If we compare, for instance, M1, M5 and M6, no notable differences in the column densities can be found: with increasing density the values are slightly higher as expected from the faster chemistry in a denser medium.
On the other hand, keeping all the other parameters unchanged, the core size may influence the abundances of c-C$_{2}$H$_{4}$O: the final c-C$_{2}$H$_{4}$O column density calculated from the output of M9 is double the value estimated for M1; the differences in the ethylene oxide abundances can vary up to one order of magnitude (see M1 and M11). This observation is explained by the fact that the larger the core the smaller its visual extinction (A$_{\rm v}$) at a certain density. The magnitude of A$_{\rm v}$ is related to the ability of the UV flux penetrating into the core and the UV flux leads in turn to the photodissociation of species; in other words, in the larger regions we will observe a greater efficiency of molecular photodissociation. 

\subsection{Sensitivity to variation in the percentage of the accreted species onto the grain surfaces}

Based on the study by Roberts et al. (2007) we introduce three non-thermal desorption mechanisms during the cold phase of our model (Phase I). Specifically, desorption resulting from: H$_{2}$ formation on grains, direct cosmic ray heating and cosmic ray-induced photodesorption. In particular, the surface formation of molecular hydrogen releases an energy of around 4.5 eV capable of inducing the thermal desorption of molecules from the grains. This is a selective process which involves mainly the most volatile species. The cosmic-ray induced photodesorption is also a selective process that accounts for the number of molecules released to the gas-phase after each cosmic ray impact. These cosmic rays are able to ionize and excite the surrounding gas resulting in the emission of a UV photon, even where the visual extinction is high enough to extinguish any UV radiation propagating in from the outside of the cloud. This internal UV flux impinges on the grain surfaces and dissociates the absorbed molecules (mainly H$_{2}$O). As for the case of H$_{2}$ formation, molecules desorb due to energy released by recombination of OH and O; this type of desorption is non-selective. All the three non-thermal desorption processes are found to be very significant in dark cloud conditions where temperatures are low enough for the thermal desorption to be insignificant. Finally, direct cosmic rays are an efficient heating source able to penetrate deeply into molecular clouds; indeed when these energetic particles impact on the grains, molecules (adsorbed on the surfaces) start to desorb because of the energy transfer from the cosmic rays to the icy mantles.

The efficiency of the non-thermal desorption of molecules depends on the percentage of the accreted species on the grain surfaces, which is in turn linked to the rate of freeze-out. The latter quantity is a free parameter in our models controlled by the product of the sticking coefficient and the average grain surface area. We vary the efficiency of freeze out (fr) in order to obtain different percentages of solid CO within a range of 75-98 \% for the hot core models and 40-98 \% for the compact ridge model. We finally investigate the effects of varying the adsorption energy that is controlled by the $\epsilon$ parameter. The latter is defined as the number of species removed each time an H$_{2}$ molecule is formed. The value of $\epsilon$ is uncertain and may be greater than 1, but, as stated by Roberts et al. (2007), at large values, the desorption will be so efficient as to prevent the build-up of any mantle material until very late times. This justifies the choice of small values for $\epsilon$, 0.01 and 0.1, in order to look at the effect of this free parameter on the desorption rates. As reported in Table 5, the column densities are slightly higher when the H$_{2}$ formation becomes the dominant non-thermal desorption mechanism ($\epsilon$ = 0.1).

\vspace{5mm}
In general, we find that all the changes made to the physical parameters (see Table 5) produce fluctuations of the estimated ethylene oxide column densities within two orders of magnitude; indeed our theoretical column densities are in the 10$^{12}$-10$^{14}$ cm$^{-2}$ range for all models. This observation may seem surprising for the case of the compact ridge models since at 80 K (the temperature adopted for these models) the co-desorption of molecules with water has not yet occurred. This explanation for the constancy of the column density of ethylene oxide may reflect the fact that non-thermal and thermal desorption provide the same contribution to the molecular abundances, but under different physical conditions. Specifically, the chemistry of denser, but smaller regions is mostly influenced by high temperature effects. However, in low density, but more extended sources, the chemistry is driven mainly by non-thermal desorption phenomena. 

\section{Results and Discussion}

\begin{figure}
 \includegraphics[width=60mm, angle=270]{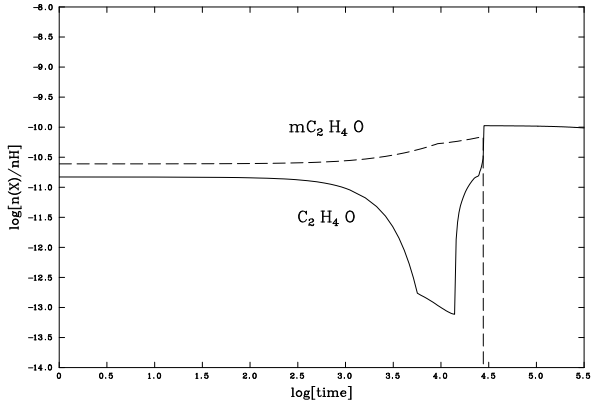}
 \caption{Fractional abundances of ethylene oxide in both gas-phase and grain surfaces as a function of time during Phase II.}
\end{figure}

\begin{table*}
 \begin{minipage}{126mm}
  \caption{Comparison between our theoretical column densities (in cm$^{-2}$) with those observed towards several high-mass star-forming regions.}
  \label{symbol}
  \begin{tabular}{@{}|ccccc}
  \hline
 & c-C$_{2}$H$_{4}$O & CH$_{3}$OH & HCOOCH$_{3}$ & CH$_{3}$CN \\
   \hline
Model & 6.2x10$^{13}$ & 2.4x10$^{17}$ & 5.6x10$^{15}$ & 2.2x10$^{16}$\\
Observations & 1.3x10$^{13}$-1.7x10$^{14}$$^{a}$ & 1.4x10$^{16}$-1.4x10$^{17}$$^{a}$ & 1.0x10$^{14}$-1.2x10$^{16}$$^{a}$ & $\ge$ 2.1x10$^{16}$$^{b}$\\
   \hline
  \end{tabular}
\footnotetext[1]{Molecular column density taken from Ikeda et al. (2001)}
\footnotetext[2]{Molecular column density taken from Wilner et al. (1994)}
 \end{minipage}
\end{table*}

Figure 2 reports the evolution of the fractional abundances of ethylene oxide in both gas-phase (solid line) and solid state (dashed line) during Phase II for Model M7 (see Table 5). At the end of the collapse, c-C$_{2}$H$_{4}$O is mostly on the grain surfaces, but its abundance drops rapidly at around 100 K when its co-desorption with water occurs (after 10$^{4}$ yrs). At the same time we observe a small growth in the gas-phase c-C$_{2}$H$_{4}$O abundances thanks to the thermal desorption of solid C$_{2}$H$_{4}$O. The amount of ethylene oxide in the gas-phase remains high at all times due to non-thermal desorption effects. 

\begin{figure}
 \includegraphics[width=60mm, angle=270]{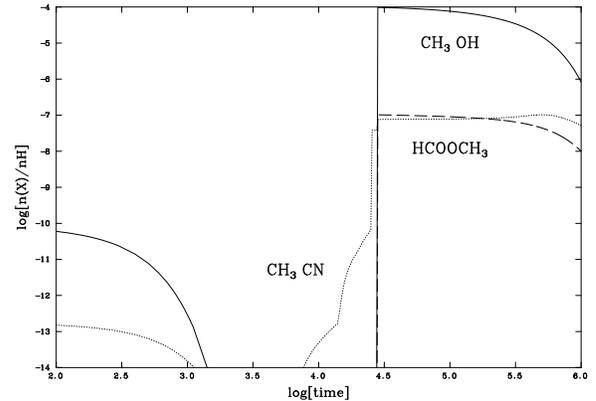}
 \caption{Theoretical fractional abundances of three typical tracers of hot cores: CH$_{3}$OH (solid line), HCOOCH$_{3}$ (dashed line) and CH$_{3}$CN (dotted line).}
\end{figure}

We now compare the theoretical column densities with the observations towards several high-mass star-forming regions where c-C$_{2}$H$_{4}$O has been detected. In general, we achieve a good agreement, within an order of magnitude, with all of the observational values for c-C$_{2}$H$_{4}$O reported in Table 3 by Ikeda et al. (2001). Referring to the range of column densities reported by these authors, we consider M7 as the model representing most of the physical characteristics of the regions listed in Ikeda et al. (2001). Results are reported in Table 6 where the calculated and observed column densities for three typical tracers for hot cores: CH$_{3}$OH, HCOOCH$_{3}$ and CH$_{3}$CN are listed. The first two species were detected by Ikeda et al. (2001), the latter molecule was investigated by Wilner et al (1994) which mapped the methyl cyanide lines in the millimeter spectrum of Orion KL region. The trends of the fractional abundances of these species as a function of time are displayed in Figure 3. The c-C$_{2}$H$_{4}$O, CH$_{3}$OH (solid line), HCOOCH$_{3}$ (dashed line) and CH$_{3}$CN (dotted line) theoretical abundances perfectly match their estimated column densities.
  
This result suggests that the reaction between atomic oxygen and ethylene on the grains is a suitable formation route for ethylene oxide.

\section{Conclusions}
We have investigated the feasibility of the reaction between oxygen ice and ethylene ice as possible route for the formation of ethylene oxide.
For this purpose, we employed a chemical model for a collapsing core by updating the rate for the formation of ethylene oxide by using the recent experimental determination of this quantity of Ward \& Price (2011). We also tested the sensitivity of ethylene oxide fractional abundances to the changes in the main physical parameters of the collapsing core. We finally reproduced the effects of non-thermal and thermal desorption of ethylene oxide during both the collapse and the warm-up phase respectively.
Our work shows a good agreement between our estimated column densities for ethylene oxide (under hot core conditions) and those observed towards several high-mass star-forming regions (Ikeda et al. 2001). In order to verify the reliability of our comparison we also looked at the theoretical and observational abundances of other molecules, such as CH$_{3}$OH, HCOOCH$_{3}$ and CH$_{3}$CN, known as good tracers of hot cores. We conclude that c-C$_{2}$H$_{4}$O can form on the grain surfaces thanks to the reaction between oxygen and ethylene.
We stressed that only 1\% of oxygen atoms depletion has to be in form of oxygen ice in order to produce ethylene oxide on the icy mantle (see Section 2); we also note that there is experimental evidence for a small quantity of ethylene being produced during the hydrogenation of acetylene to form ethane; therefore it is robust to assume that 1$\%$ of acetylene is hydrogenated to solid ethylene as an initial condition of our model. This amount of ethylene is already enough for the observed c-C$_{2}$H$_{4}$O column densities to be reproduced by our theoretical results. Further experimental studies are perhaps needed to extend our study to other cyclic species that have been detected in the ISM, although the complexity of larger molecules may prevent their isomeric speciation as is the case for the recent laboratory study of the formation of C$_{3}$H$_{6}$O.
 
\section{Acknowledgments}
The research leading to these results has received funding from the [European Community's] Seventh Framework Program [FP7/2007-2013] under grant agreement n$^{\circ}$ 238258.

The authors thank the anonymous referee for useful suggestions that greatly improved the manuscript.

\label{lastpage}
\end{document}